\documentclass[prl,twocolumn,showpacs,superscriptaddress,aps,amsfonts]{revtex4}
\usepackage{graphicx,graphics,color,epsfig}
\usepackage{bm}
\usepackage{amsmath}
\usepackage{amssymb}
\usepackage{epstopdf}

\begin{document}
\bibliographystyle{prsty}
\title{
Role of bosonic modes in the mechanism of high
           temperature Bi$_{2}$Sr$_{2}$CaCu$_{2}$O$_{8+\delta}$
           superconductors using ultrafast optical techniques}
\author{Jian-Xin Zhu}
\email{jxzhu@lanl.gov}
\homepage{http://theory.lanl.gov}
\affiliation{Los Alamos National Laboratory, Los Alamos, NM 87545,
USA}
\author{Elbert E. M. Chia}
\email{ElbertChia@ntu.edu.sg}
\affiliation{Division of Physics and Applied Physics, School of
Physical and Mathematical Sciences, Nanyang Technological
University, Singapore 637371, Singapore}
\affiliation{Los Alamos
National Laboratory, Los Alamos, NM 87545, USA}
\author{T. Tamegai}
\affiliation{Department of Applied Physics, The University of Tokyo,
Hongo, Bunkyo-ku, Tokyop 113-8656, Japan}
\author{H. Eisaki}
\affiliation{AIST Tsukuba Central 2, 1-1-1 Umenzono, Tsukuba,
Ibaraki 305-8568, Japan}
\author{Kyu-Hwan Oh}
\affiliation{Department of Physics, Pohang University of
Science and Technology, Pohang 790-784, Republic of Korea}
\author{S.-I. Lee}
\affiliation{National Creative Research Initiative Center for Superconductivity and
Department of Physics, Seoul 121-742, Republic of Korea}
\author{A. J. Taylor}
\affiliation{Los Alamos National Laboratory, Los Alamos, NM 87545,
USA}
\date{\today}

\begin{abstract}
Using ultrafast optical techniques, we probe the hole-doping
dependence of the electron-boson coupling constant $\lambda$ in
Bi$_{2}$Sr$_{2}$CaCu$_{2}$O$_{8+\delta}$. In the overdoped region,
we observe a correlation between ($\lambda$) and the superconducting
transition temperature $T_{c}$. Upon performing the McMillan
analysis, however, we find that $\lambda$ is too small to explain
the high $T_{c}$'s, and that the Coulomb pseudopotential
$\mu^{\ast}$ is negative. Our analysis therefore reveals two
components in the mechanism of high-$T_{c}$ superconductivity --- a
dominant pre-existing pairing interaction, together with a weaker
electron-phonon interaction that fine-tunes $T_{c}$.
\end{abstract}


\maketitle 

Despite many advances in understanding copper-oxide
high-temperature superconductors, there still exists no universally
accepted mechanism. Determining the nature of interaction
responsible for the Cooper-pair formation remains one of the grand
challenges in modern condensed matter physics. The most probable
candidates are lattice vibrations (phonons)
\cite{McQueeney99,Lanzara01}, spin fluctuation modes
\cite{Mignod91,Norman97}, and pairing without invoking glue
\cite{Anderson07}. For conventional superconductors, structure in
the electron tunneling $dI/dV$ characteristics established
unambiguously that the attractive pairing interaction was mediated
by phonons \cite{McMillan69}. For high transition temperature
($T_{c}$) superconductors, structure in $dI/dV$ has also been found
in many tunneling measurements \cite{Kirtley07}. More recent
scanning tunneling microscopy (STM) experiments revealed an oxygen
lattice vibration mode whose energy is anticorrelated with the local
gap value on hole-doped Bi$_{2}$Sr$_{2}$CaCu$_{2}$O$_{8+\delta}$
(Bi-2212) \cite{Lee06} while a bosonic mode of electronic origin was
found in the electron-doped Pr$_{0.88}$LaCe$_{0.12}$CuO$_{4}$
\cite{Niestemski07}. Together with salient features observed in
angle-resolved photoemission spectroscopy (ARPES)
\cite{Lanzara01,Cuk04,Gweon04}, these new results raise the
fundamental question of whether the bosonic modes are a pairing glue
\cite{Balatsky06} or a signature of an inelastic tunneling channel
\cite{Pilgram06}. Here, we report a systematic time-resolved
pump-probe study on Bi-2212 at various doping levels. It reveals a
positive correlation between the quasiparticle relaxation rate and
$T_{c}$ as doping is varied from the optimal toward the overdoped
regime, indicating that phonons play a role in the mechanism of
high-$T_{c}$ superconductivity. Our analysis, based on McMillan-type
strong coupling theory, shows that (i) the electron-phonon coupling
is not sufficiently large to account for the large $T_{c}$'s of the
cuprates, and (ii) the Coulomb pseudopotential is necessarily
\textit{negative}, indicating that a dominant pre-existing pairing
interaction is necessary to glue the electrons into pairs with such
high $T_{c}$'s. Candidates of such a pre-existing pairing
interaction are electronic coupling to a bosonic
mode of electronic origin, or a mechanism without mediators.

The role of the electron-boson interaction (EBI) in high-$T_{c}$
superconductors were studied by different techniques. For example,
inelastic neutron scattering tracks the changes in boson energies or
dispersions upon entering the superconducting state. ARPES and
time-integrated optical spectroscopy~\cite{Carbotte99,Hwang04}
measure the effects of EBI on electronic self-energies, and planar
junction experiments determine the energy of the bosonic 
mode~\cite{Zasadzinski01}. STM experiments measure the local density of
states through the local differential tunneling conductance, where
the characteristic boson mode energy is estimated from the peak
position in $d^2I/dV^2$~\cite{Lee06}. However, it cannot tell us directly the
strength of the electron-boson coupling because all energy is
encoded in the electron self-energy itself. Complementary to the
above techniques, ultrafast spectroscopy -- a temporally-resolved
technique, has been used in probing the relaxation dynamics of
photoexcited quasiparticles in correlated electron 
systems~\cite{Averitt02,Demsar06}. Its unique contribution lies in its
ability to extract the value of the electron-boson coupling constant
($\lambda$) directly, without the need to perform complicated
inversion algorithms. This procedure has been experimentally
verified on the conventional superconductors~\cite{Brorson90}.
Performing time-resolved pump-probe measurements on the
\textit{same} family of cuprates will allow us to determine (i) the
magnitude of $\lambda$, and (ii) whether $\lambda$ has any
correlation with doping. These will yield crucial information to the
role of electron-boson interaction in the mechanism of high-$T_{c}$
superconductivity.

The family of the two-layer cuprate Bi-2212 has been in recent years
the most intensively studied class of high-$T_{c}$ superconductors,
due to their (a) extreme cleavability, (b) containing only CuO$_{2}$
planes and not chains, and (c) the possibility of growing samples
with a larger range of $T_{c}$'s (compared to other cuprates).
Single crystals of Bi-2212 were obtained from two groups (Tokyo and
AIST) grown by the floating zone method with doping controlled by
oxygen depletion, yielding values of $T_{c}$ (determined by
magnetization data) that depends on the hole doping level spanning
from the underdoped to the overdoped regime. Due to difficulties in
growing high-purity underdoped samples, we present data on only one
underdoped sample, and six optimally-doped to overdoped samples.

\begin{figure}
\centering \includegraphics[width=8cm,clip]{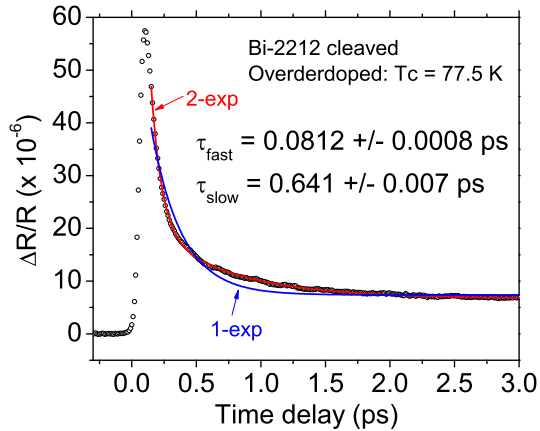}
\caption{(Color) Photoinduced transient reflection $\Delta R/R$ versus time
delay between pump and probe pulses, of an overdoped Bi-2212 single
crystal ($T_{c}$ = 90.5~K). (o): Experimental data. Blue line:
one-exponential fit. Red line: two-exponential fit.}
\label{fig:Bi2212Fig1}
\end{figure}

In our experiments an 80-MHz repetition rate Ti:sapphire laser
produces 45-femtosecond (fs) pulses at approximately 800~nm (1.5~eV)
as the source of both pump and probe optical pulses. The pump and
probe pulses were cross-polarized, with a pump spot diameter of
$\sim$60 $\mu$m and probe spot diameter of $\sim$30 $\mu$m. The
reflected probe beam was focused onto an avalanche photodiode
detector. The pump beam was modulated at 1~MHz with an
acoustic-optical modulator to minimize noise. The experiments were
performed with an average pump power of 3~mW, giving a pump fluence
of $\sim$0.1~J/cm$^{2}$ and a photoexcited quasiparticle density of
0.02/unit cell, showing that the system is in the weak perturbation
limit. The probe intensity was $\sim$10 times lower. Resolution is
at least 1 part in 10$^{6}$. The fitted values of $\tau$  have a
typical error of $\pm$1 $\%$. All measurements are done at room
temperature. At this temperature, the electron subsystem reaches a
local equilibrium within 20~fs, shorter than our pulse width. Hence
the observed quasiparticle relaxation occurs mostly through
electron-lattice coupling. Figure~\ref{fig:Bi2212Fig1} shows the
time dependence of the photoinduced signal of a typical overdoped
Bi-2212 sample. The time evolution of the photoinduced reflection
$\Delta R/R$ first shows a rapid rise time (of the order of the pump
pulse duration) followed by a subsequent decay. 
As shown in Fig.~\ref{fig:Bi2212Fig1}, the data can be fit better by two
exponentials (red line) than a single exponential (blue line). It
indicates the quasiparticle relaxation has two components: $\Delta
R/R = A + B \exp(-t/\tau_{fast}) + C \exp(-t/ \tau_{slow})$. The fast
component $\tau_{fast}$ is of the order of 100~fs while the slow
component $\tau_{slow}$ is of the order of 650~fs in the
optimamly-doped to overdoped regimes. Since observation of
spin-fluctuation modes in the overdoped Bi-2212 samples has rarely
been reported, we ascribe the quasiparticle relaxation in this
regime to electron-lattice coupling, \textit{not} coupling between
electrons and spin fluctuations --- this is consistent with STM data
on the same family of cuprates~\cite{Lee06}. Recently a
time-resolved photoelectron spectroscopy measurement has been
carried out on an optimally doped Bi-2212 sample, where a similar
two-stage cooling dynamics was observed at room temperature
\cite{Perfetti07}. Similarly, we interpret the relaxation process as
the electrons first transferring energy to the phonons which are
more strongly coupled at a characteristic time $\tau_{fast}$ and
then continue cooling down via the energy dissipation of these hot
phonons by the means of anharmonic decay at a characteristic time
$\tau_{slow}$.

\begin{figure}[t]
\centering 
\includegraphics[width=8cm,clip]{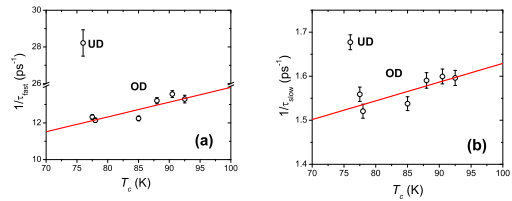}
\caption{(Color) Doping dependence of the (a) fast relaxation rate
($1/\tau_{fast}$) and (b) slow relaxation rate ($1/\tau_{slow}$). UD
= underdoped sample. OD = overdoped samples. Red lines = best-fit
straight line through the OD data points.} \label{fig:Bi2212Fig2}
\end{figure}

\begin{figure} \centering \includegraphics[width=8cm,clip]{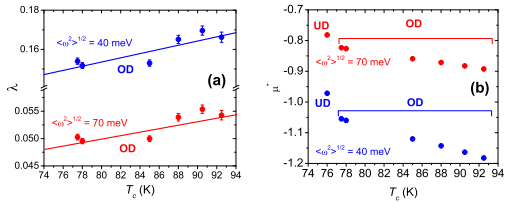}
\caption{(Color) Doping dependence of the (a) electron-phonon coupling
constant $\lambda$ (OPT and OD samples only), and (b) effective
Coulomb pseudopotential $\mu^{\ast}$, for $\sqrt{\left\langle
\omega^{2} \right\rangle}$ = 40~meV (black circles) and 70~meV (red
circles). The read and blue lines in (a) are best-fit straight lines
through the OPT and OD points.} \label{fig:Bi2212Fig3}
\end{figure}

Figure~\ref{fig:Bi2212Fig2} shows the doping dependence of the
relaxation rate, $1/\tau_{fast}$ and $1/\tau_{slow}$. In the
optimally-doped to overdoped regimes, both relaxation rates are
correlated with $T_{c}$, though weakly. Moreover, the relaxation
rate in the underdoped regime deviates systematically from those in
the optimally-doped and overdoped regimes. This significant
deviation suggests that the quasiparticle relaxation in the
underdoped regime has, in addition to the electron-phonon
interaction, another mechanism, an obvious candidate being the
interaction between quasiparticles and spin-fluctuations, due to the
superconducting dome being in close proximity to an otherwise
undoped antiferromagnetic insulator. The existence of a correlation
between $1/\tau$ and $T_{c}$ suggests that phonons play a role in
the mechanism of high-$T_{c}$ superconductivity in the cuprates.

Next, we perform a quantitative analysis by assuming that the
quasiparticle relaxation throughout the entire doping regime is due
to the electron-phonon interaction. Since, as already mentioned, the
transfer of electron energy first occurs through selected modes that
are most strongly coupled to electrons, we use $\tau_{fast}$ to
estimate the electron-phonon coupling strength $\lambda$. These
strongly coupled phonon modes should be the most relevant in
discussing the possible phonon-mediated superconductivity. We
consider the out-of-plane out-of-phase oxygen buckling $B_{1g}$
phonon and the half-breathing in-plane copper-oxygen bond stretching
phonon with energies of approximately $\Omega$ = 40 and 70 meV,
respectively. These two types of phonon modes are suggested to be
responsible for the dispersion anomalies at the antinodal
\cite{Cuk04} and nodal \cite{Lanzara01} directions, and reveal
strong line-shape renormalizations with doping and temperature in
Raman and neutron scattering measurements
\cite{McQueeney99,Pyka93,Reznik95,McQueeney01,Sugai03}. Though
cuprate samples like Bi-2212 are inhomogeneous both in energy gap
and characteristic boson frequency, the spatial average of mode
frequency is doping \textit{independent}. Therefore, we assume
$\Omega$ is constant throughout the entire doping regime. $\tau$ is
related to $\lambda$ by the Allen relation \cite{Allen87}

\begin{equation}
\frac{1}{\tau}=\frac{3\hbar \lambda \left\langle \omega^{2}
\right\rangle}{\pi k_{B}T_{e}}, \label{eqn:Allen}
\end{equation} where $T_{e}$ is the electronic temperature (estimated to be 340~K), $\lambda \left\langle
\omega^{2} \right\rangle$ is the second moment of the effective
electron-phonon coupling strength, $\alpha^{2}F(\omega)$. We use the
Einstein model for phonons such that $\left\langle \omega^{2}
\right\rangle = \Omega^{2}$ and $\left\langle \omega \right\rangle =
\Omega$. Figure~\ref{fig:Bi2212Fig3}(a) shows the calculated values of
$\lambda$ from Eq.~(\ref{eqn:Allen}), as a function of $T_{c}$. Since
$1/\tau \propto \lambda$ , $\lambda$ also correlates with $T_{c}$.
Within the strong-coupling theory, the McMillan
\cite{McMillan69,McMillan68} formula for the superconducting
transition temperature in a $d$-wave superconductor is found to be
\cite{Zhu07}:

\begin{equation}
T_{c} = \omega_{0} \text{ exp} \left
[-\frac{2(1+\lambda)}{\lambda-\mu^{\ast}(1+\lambda \left \langle
\omega \right \rangle)/2 \omega_{0}} \right ]. \label{eqn:Tc}
\end{equation} Here $\mu^{\ast}$  is the Coulomb pseudopotential; it is a
renormalized quantity and can be very different from the original
bare Coulomb repulsion. In this formula, the factor 2 accounts for
the $d$-wave nature of superconducting order parameter while the
renormalization factor ($1+\lambda$)  arises from the $s$-wave
channel of the electron-phonon coupling. Since the average bosonic
mode frequency $\omega_{0}$ remains unchanged with doping
\cite{Lee06}, the variation of $T_{c}$ across the superconducting
dome in Bi-2212 is due solely to the interplay between $\lambda$ and
$\mu^{\ast}$. With the given coupling strength $\lambda$ and
$\omega_{0} \approx \Omega$, and the experimentally measured
$T_{c}$, one imposes a stringent constraint on $\mu^{\ast}$ through
Eq.~(\ref{eqn:Tc}). Figure \ref{fig:Bi2212Fig3}(b) shows the calculated
values of $\mu^{\ast}$ for both phonon modes. The
\textit{negativity} of $\mu^{\ast}$ presents two important
implications: (i) the electron-phonon interaction \textit{alone} is
not strong enough to cause Bi-2212 to be superconducting at the
measured high $T_{c}$, and (ii) a pre-existing (attractive) pairing
interaction is necessary, especially when no bosonic modes of
electronic origin (such as spin resonance modes) exist, which is
true at least in the deeply overdoped region where they are neither
expected nor observed. Figure~\ref{fig:Bi2212Fig3}(b) shows a clear
positive correlation between $\mu^{\ast}$ and $T_{c}$, with the
magnitude of $\mu^{\ast}$ reaching a maximum at the optimal doping,
which shows that a large pre-existing pairing interaction is needed
to produce a large $T_{c}$. Our data on an almost optimally-doped
three-layered cuprate Tl$_{2}$Ba$_{2}$Ca$_{2}$Cu$_{3}$O$_{y}$, with
a higher $T_{c}$ of 115~K (not shown here), revealed even smaller
values of $1/\tau_{fast}$ and $1/\tau_{slow}$ than the corresponding
values of all our Bi-2212 samples, supporting our assertion that
electron-phonon coupling \textit{alone} cannot be the pairing
mechanism for high-$T_{c}$ superconductivity.

\begin{figure} \centering \includegraphics[width=8cm]{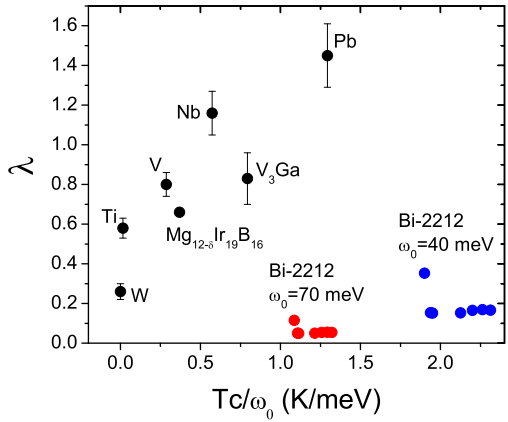}
\caption{(Color) $\lambda$ versus $T_{c}/\omega_{0}$ for different classes
of superconductors. Solid black circles = coventional
superconductors \cite{Brorson90,Mu07}. Solid red circles = Bi-2212
for $\omega_{0}=\sqrt{\left\langle \omega^{2} \right\rangle}$ =
70~meV. Solid blue circles = Bi-2212 for $\sqrt{\left\langle
\omega^{2} \right\rangle}$ = 40~meV.} \label{fig:Bi2212Fig4}
\end{figure}

In Figure~\ref{fig:Bi2212Fig4} we plot $\lambda$ versus
$T_{c}/\omega_{0}$, combining the data for Bi-2212 with some
conventional ($s$-wave) superconductors, also obtained from
pump-probe measurements \cite{Brorson90}, as well as the recently
discovered nodeless non-centrosymmetric superconductor
Mg$_{12-\delta}$Ir$_{19}$B$_{16}$~\cite{Mu07}. Compared to these
conventional superconductors, we notice that data points for Bi-2212
(i) do not follow the same trend, (ii) show a much weaker
correlation between $\lambda$ and $T_{c}/\omega_{0}$, with a much
smaller slope, (iii) have smaller values of $\lambda$, and (iv) have
larger values of $T_{c}/\omega_{0}$. If the McMillan formula is
valid for Bi-2212, then, for a positive $\mu^{\ast}$, a large
$T_{c}/\omega_{0}$ should imply a large $\lambda$, i.e. the Bi-2212
data points should lie on the same trend as the conventional
superconductors. However, the values of $\lambda$, directly obtained
from $1/\tau$, are too small. For Eq.~\ref{eqn:Tc} to still hold, we
therefore need a \textit{negative} $\mu^{\ast}$, i.e. we return to
the same conclusion of the need for a pre-existing pairing
interaction.

The central conclusion of this work, is that \textit{both} a
pre-existing pairing interaction \textit{and} the electron-phonon
interaction play a role in the mechanism of high-$T_{c}$
superconductivity. The pre-existing pairing interaction plays a
dominant role, while the electron-phonon coupling, in cooperation
with the pre-existing interaction, merely fine-tunes the transition
temperature as evidenced by its weak positive correlation with
$T_{c}$. Our finding therefore echoes the hypothesis by Anderson
\cite{Anderson07} that one cannot neglect the ultimate importance of
strong correlation effects to explain the mechanism for high-$T_{c}$
superconductivity in the cuprates, where the Coulomb repulsion is
comparable or even larger than the relevant $d$-orbital bandwidth.
It also suggests that, though we can use a convenient effective
theory of electronic coupling to bosonic modes to understand many
interesting signatures observed in ARPES and tunneling experiments
on cuprates, such a theory does not reveal the underpinning
mechanism for superconductivity. Theoretically, the
spin-fluctuation-mediated pairing mechanism is of electronic origin
and should come from the same strong electronic correlation. Recent
dynamical cluster calculations \cite{Maier06} have indeed shown that
the pairing mechanism in the doped two-dimensional Hubbard model is
mediated by the exchange of $S$=1 particle-hole spin fluctuations.
So far, whether the extracted coupling between quasiparticles and
spin fluctuations is a dominant mechanism or a secondary effect
remains hotly debated in a strong electronic correlation model for
superconductivity. Experimentally, if there exists a dominant
electron-boson (specifically spin fluctuation mode) coupling, it
will be very interesting to characterize directly the strength of
this coupling of electronic origin.

We acknowledge useful discussions with C. Panagopoulos, A. V.
Balatsky, M. Graf, S. A. Trugman, and D. Mihailovic. This work was
carried out under the auspices of the National Nuclear Security
Administration of the U.S. DOE at LANL under Contract No. DE-AC52-06NA25396, 
the U.S. DOE Office of Science,  the LANL LDRD Program, and the Singapore
Ministry of Education Academic Research Fund Tier 1 under Grant
number RG41-07.

%

\end{document}